\documentclass[12pt]{article}
\pdfoutput=1
\usepackage{epsf,amsfonts,amssymb,epsfig,amsmath,mathtools,graphics,slashed}
\usepackage{hyperref,hep,graphicx,subfig}
\usepackage{dsfont}
\usepackage{color}

\makeatletter

\newcommand{\Rmnum}[1]{\expandafter\@slowromancap\romannumeral #1@}
\makeatother

\newcommand{\nn}{\notag \\}

\begin{document}

\makeatletter
\renewcommand{\theequation}{\thesection.\arabic{equation}}
\@addtoreset{equation}{section}
\makeatother

\baselineskip 18pt

\begin{titlepage}

\vfill

\begin{flushright}
Imperial/TP/2015/JG/02\\
\end{flushright}

\vfill

\begin{center}
   \baselineskip=16pt
   {\Large\bf A new phase for the anisotropic\\ N=4 super Yang-Mills plasma}
  \vskip 1.5cm
  \vskip 1.5cm
      Elliot Banks and Jerome P. Gauntlett\\
   \vskip .6cm
         \vskip .6cm
      \begin{small}
      \textit{Blackett Laboratory, 
        Imperial College\\ London, SW7 2AZ, U.K.}
        \end{small}\\

\end{center}

\vfill

\begin{center}
\textbf{Abstract}
\end{center}

\begin{quote}
Black hole solutions of type IIB supergravity have been previously constructed that describe the
N=4 supersymmetric Yang-Mills plasma with an anisotropic spatial deformation. The zero temperature limit of these black holes approach a Lifshitz-like scaling solution in the infrared. 
We show that these black holes
become unstable at low temperature and we construct a new class of black hole solutions 
which are thermodynamically 
preferred. The phase transition is third order and incorporates a spontaneous breaking
of the $SO(6)$ global symmetry down to $SO(4)\times SO(2)$. The critical exponents for the phase transition are given by $(\alpha,\beta,\gamma,\delta)=(-1,1,1,2)$ 
which differ from the standard mean-field exponents usually seen in holography. 
At low temperatures the black holes approach a novel kind of scaling behaviour in the far IR
with spatial anisotropy and hyperscaling violation. We show that the new ground states are thermal insulators in the direction of the anisotropy.
\end{quote}

\vfill

\end{titlepage}
\setcounter{equation}{0}

%\tableofcontents

\section{Introduction}
Strongly coupled conformal field theories that have been deformed by spatially dependent sources in flat spacetime can be studied by constructing novel black hole solutions using the AdS/CFT correspondence (e.g.
\cite{Azeyanagi:2009pr,Mateos:2011ix,Mateos:2011tv,Hartnoll:2012rj,Horowitz:2012ky,Horowitz:2012gs,Donos:2012js,Chesler:2013qla,Ling:2013nxa,Donos:2013eha,
Andrade:2013gsa,Balasubramanian:2013yqa,Donos:2014uba,Gouteraux:2014hca,Cheng:2014qia,Donos:2014cya,Jain:2014vka,Donos:2014oha,Donos:2014yya,Kim:2014bza,
Davison:2014lua,Donos:2014gya,Andrade:2014xca,Blake:2015ina,Kiritsis:2015oxa}).
Such studies are interesting for a number of reasons. Without such sources the translational symmetry implies that momentum cannot dissipate and this leads to 
non-physical delta-function responses in, for example, the thermal and electric conductivity of the system. Spatially dependent sources provides
a mechanism for momentum to dissipate and this leads to finite DC responses. The spatially dependent sources also provide a useful tool to search for novel
holographic ground states which can appear in the far IR; 
insulators, coherent metals and incoherent metals have been realised in this way, as well as transitions between them \cite{Donos:2012js,Donos:2013eha,Donos:2014uba,Gouteraux:2014hca,Kiritsis:2015oxa}.
A more specific motivation derives from the properties of the quark gluon plasma observed in heavy ion collisions. In particular,
the plasma appears to have regimes where it is described by a strongly coupled and spatially anisotropic fluid \cite{Shuryak:2003xe,Shuryak:2004cy}.

An interesting framework for analysing spatial anisotropy in N=4 super Yang-Mills (SYM) theory was initiated in \cite{Azeyanagi:2009pr} and then further
developed in \cite{Mateos:2011tv,Mateos:2011ix}.
Specifically, black hole solutions of type IIB supergravity were constructed in
\cite{Mateos:2011tv,Mateos:2011ix} that asymptotically approach
$AdS_5$ at the UV boundary with the type IIB axion having a linear dependence on one of the three spatial coordinates. 
The linear axion source is associated with a distribution of D7-branes that intersect D3-branes in two of the spatial directions and
is smeared in the third. At low temperatures these black holes approach a $T=0$ solution, constructed in \cite{Azeyanagi:2009pr}, 
which becomes a Lifshitz-like scaling solution in the far IR.

It is natural to interpret this scaling solution as the $T=0$ ground state of the anisotropically deformed N=4 SYM theory. Here, however, we will show
that the black holes of \cite{Mateos:2011tv,Mateos:2011ix} are unstable at low temperatures and there is a phase transition
which spontaneously breaks the global $SO(6)$ symmetry down to $SO(4)\times SO(2)$. The origin of this
instability was already noticed in \cite{Azeyanagi:2009pr}. In particular,
by analysing the Kaluza-Klein spectrum
of the five-sphere it was found that there are scalar modes, transforming in the ${\bf 20}'$ of $SO(6)$, 
which saturate the BF bound in $AdS_5$ background but violate an analogous bound in the Lifshitz-like background. This suggests that the
Lifshitz-like scaling solution is unstable.  It is natural to suspect that such an instability is also present for the $T=0$ solution that interpolates between
$AdS_5$ in the UV and the Lifshitz-like solution in the IR. By continuity one then expects that the finite temperature black hole solutions should
become unstable at some finite temperature.

Here we will show that these expectations are realised. We find that the black holes become unstable at some critical temperature $T_c$
with two new branches of black holes appearing. One of these branches seems to be a non-physical branch of ``exotic hairy black holes" 
\cite{Buchel:2009ge,Donos:2011ut}.
Specifically, it seems that these black holes only exist for temperatures $T\ge T_c$ and are never thermodynamically preferred. The other branch exists for
$T\le T_c$ and is thermodynamically preferred.
The phase transition is continuous and, somewhat surprisingly, third order. 
Furthermore, we calculate the critical exponents of the phase transition finding
 $(\alpha,\beta,\gamma,\delta)=(-1,1,1,2)$ rather than the
standard mean-field values $(\alpha,\beta,\gamma,\delta)=(0,1/2,1,3)$ associated
with most holographic phase transitions.
The critical exponents we find 
have been previously realised in bottom-up models of holographic superconductors with a Lagrangian containing 
cubic terms in the modulus of the complex scalar field \cite{Franco:2009yz,Franco:2009if,Herzog:2010vz,Aprile:2010yb}. 
While such terms are a bit unnatural for a 
complex scalar field they are natural for a neutral scalar field, provided the potential does not have any discrete symmetry. The critical 
exponents that we find can, in a certain sense, be realised by a Landau-Ginzburg (LG)
model with a scalar order parameter and cubic term in the free energy. While such cubic terms in LG models are associated with first order phase transitions, our holographic transition appears
to be continuous and third order.

We construct our new solutions using a consistent KK truncation of type IIB supergravity on $S^5$ that keeps the $D=5$ metric coupled to the axion and dilaton, as in \cite{Mateos:2011tv,Mateos:2011ix}, and in addition keeps an extra single neutral scalar field. Any solution of the $D=5$ theory gives rise
to an exact solution of type IIB supergravity. The potential for the neutral scalar field in the $D=5$ theory does not have any
discrete symmetry and this is associated with the non-standard values of the critical exponents that we just discussed. The fact that the
phase transition spontaneously breaks
$SO(6)$ to $SO(4)\times SO(2)$ is only apparent after
uplifting to type IIB.

We construct the new branch of black hole solutions down to low temperatures and elucidate the $T=0$ behaviour.
Similar to \cite{Donos:2012js} and unlike many holographic studies, the IR part of the geometry at $T=0$ does not approach a scaling solution
of the equations of motion, but instead approaches the leading terms of an expansion, which eventually approaches $AdS_5$ in the UV.
The leading terms of this IR expansion are similar to the hyperscaling
violation solutions \cite{Charmousis:2010zz,Ogawa:2011bz,Huijse:2011ef}
but with anisotropic scaling in the spatial direction rather than the time direction.
We calculate the thermal conductivity of the black holes, essentially importing the results of \cite{Donos:2014cya}. We show that
the scaling behaviour  implies that at low temperatures the system is a thermal insulator with $\kappa\sim T^{10/3}$.

The remainder of the paper is organised as follows. In section \ref{action}, we present the $D=5$ top-down model and show how it
is arises from a KK reduction of type IIB supergravity on $S^5$. 
The construction of the black holes and a study of their properties is contained in section \ref{back-reacted}.
We conclude in section \ref{concl} and we have two appendices. Appendix A contains
some technical results concerning a Smarr formula, while appendix B discussed
how the critical exponents that we find can be extracted, in a certain sense, 
from a Landau-Ginzburg type analysis.

\section{The top-down model} \label{action}
We will consider a $D=5$ gravity theory coupled to three scalar fields, the axion and dilaton, $\phi$ and $\chi$, as in \cite{Mateos:2011ix,Mateos:2011tv}, 
and an additional scalar $X$.
The bulk action is given by:
\begin{align}
\label{5dAction}\begin{split}
S%_{bulk} 
= &\int
d^5x\sqrt{-g}\left(R-3X^{-2}(\partial X)^2+4(X^2+2X^{-1}) -\frac{1}{2}(\partial\phi)^2-\frac{1}{2}e^{2\phi}(\partial\chi)^2\right)\,,
 %\\&+ \frac{1}{2\kappa^2}\int_{\partial\mathcal{M}}\sqrt{-\gamma}2K,
\end{split}\end{align}
where for simplicity of presentation we have set $16\pi G=1$.
The corresponding equations of motion are given by:
\begin{align}\label{5dEEquation}
\nabla^2\phi &= e^{2\phi}(\partial \chi)^2\,,
\nn
\nabla_\mu\left(e^{2\phi}\nabla^\mu\chi\right) &= 0\,,
\nn 
\nabla_\mu\left(X^{-1}\nabla^\mu X\right) &= -\frac{4}{3}(X^2-X^{-1})\,,\nn
 R_{\mu\nu} = \  3X^{-2}\partial_\mu X\partial_\nu X&-\frac{4}{3}(X^2+2X^{-1})g_{\mu\nu}
%\\ &
+\frac{1}{2}\partial_\mu \phi \partial_\nu \phi  + \frac{1}{2}e^{2\phi}\partial_\mu \chi \partial_\nu \chi\,.
\end{align}

This top-down model arises as a consistent truncation of the Kaluza-Klein (KK) reduction of type IIB supergravity on a five-sphere.
That is, any solution to the equations of motion \eqref{5dEEquation} gives rise to an exact solution of type IIB supergravity with
$D=10$ metric and self-dual five-form given by:
\begin{align}\label{KKansatz}
d{s}^2_{10} =\  &\bar\Delta^{1/2}ds_5^2 +  X \bar\Delta^{1/2}d\xi^2 + X^2\bar\Delta^{-1/2}\sin^2\xi d\tau^2
%\\  &
+\bar\Delta^{-1/2}X^{-1}\cos^2\xi d\Omega_3\,,
\nn
F_{(5)} = \ &2U\text{vol}_5 + 3\sin\xi \cos\xi X^{-1}*_5dX\wedge d\xi\nn
&+\bar\Delta^{-2}\sin\xi\cos^3\xi (2Ud\xi - 3\sin\xi\cos\xi X^{-2} dX )\wedge d\tau\wedge \text{vol}_3\,,
%\\ \hat{\phi} = \ &\phi, \qquad \hat{\chi} = \chi
\end{align}
where $ds^2_5$ and $\text{vol}_5$ are the $D=5$ metric and volume form, respectively, $d\Omega_3$ and
$\text{vol}_3$ are the metric and volume form on a round three-sphere, respectively, and
\begin{align}\label{delbar}
%c &= \cos\xi, \qquad s = \sin\xi\nn 
\bar\Delta &= X^{-2}\sin^2\xi + X\cos^2\xi,\qquad
 U = X^2\cos^2\xi + X^{-1}\sin^2\xi +X^{-1}\,.
\end{align}
The $D=10$ dilaton and axion are the same as the $D=5$ scalar fields $\phi$ and $\chi$, respectively, and the $D=10$ three-forms are both zero. When $X\ne 0$ this class of $D=10$ metric and five-form has the $SO(6)$ symmetry of the round five-sphere reduced to
$SO(4)\times SO(2)$, with the first factor acting on the round $S^3$ and the second acting on the circle parametrised by $\tau$. 

That this is a consistent truncation can be established using the results of \cite{Lu:1999bw}. Indeed it was shown in
\cite{Lu:1999bw} that there is a consistent KK truncation of type IIB supergravity on a five-sphere to Romans $D=5$
$SU(2)\times U(1)$ gauged supergravity, whose bosonic fields consist of a $D=5$ metric, a scalar $X$,  $SU(2)\times U(1)$ gauge-fields
and two two-forms. This truncation can simply be extended to include the $D=10$ axion and dilaton and
we can then truncate away the gauge-fields and the two-form to obtain our model. 

Notice that the unit radius $AdS_5$ vacuum solution to the equations of motion \eqref{5dEEquation}
has $\chi=0$, $X=1$ with constant $\phi$, and uplifts to the standard $AdS_5\times S^5$ solution of type IIB.
Around this vacuum solution, perturbations of the fields $\phi,\chi$ are massless and are associated with marginal operators in $N=4$ SYM
theory with scaling dimension $\Delta=4$. Perturbations of $X$ have $m^2=-4$, which saturates the BF bound, and is associated with an operator ${\mathcal O}_\psi$ with dimension $\Delta=2$. This operator is part of a multiplet, transforming in the ${\bf 20}'$
of $SO(6)$ which is dual to operators in N=4 SYM constructed from the six adjoint scalar fields $\phi^I$ of the form $Tr(\phi^I\phi^J)-trace$. When ${\mathcal O}_\psi$ acquires an expectation value spontaneously, as it will in our solutions, it breaks the $SO(6)$ 
global $R$-symmetry down to $SO(4)\times SO(2)$.

Notice that setting $X=1$, which is a further consistent truncation, we have $\bar\Delta=1$, $U=2$, from \eqref{delbar}, and we recover the
$D=5$ model that was studied in  \cite{Mateos:2011ix,Mateos:2011tv}. 
In particular the metric on the five-sphere in \eqref{KKansatz} becomes
the round metric.
Thus, our model extends the top-down model studied in \cite{Mateos:2011ix,Mateos:2011tv}
to include one extra scalar field, $X$, which saturates the BF bound. 
It is sometimes convenient to consider a canonically normalised scalar field, $\psi$, instead of $X$, defined by
\begin{align}\label{defpsi}
%X\equiv e^{-\frac{\psi}{\sqrt 6}}\\
X\equiv e^{-{\psi}/{\sqrt 6}}\, ,
\end{align}
in terms of which the action reads
\begin{align}
\label{5dAction2}\begin{split}
S%_{bulk} 
= &\int
d^5x\sqrt{-g}\left(R-\frac{1}{2}(\partial\psi)^2+4(e^{-{2\psi}/{\sqrt 6}}+2 e^{{\psi}/{\sqrt 6}})
-\frac{1}{2}(\partial\phi)^2-\frac{1}{2}e^{2\phi}(\partial\chi)^2\right)\,.
 %\\&+ \frac{1}{2\kappa^2}\int_{\partial\mathcal{M}}\sqrt{-\gamma}2K,
\end{split}\end{align}

\subsection{Brief review of previous work} \label{linear}
The anisotropic solutions constructed in \cite{Azeyanagi:2009pr,Mateos:2011ix,Mateos:2011tv}, with the axion linear in one of the spatial coordinates,
all lie within the ansatz
\begin{align}\label{ansatz1}
ds^2 &= \frac{e^{-\frac{1}{2}\phi}}{u^2}\left(-\mathcal{FB}dt^2+\frac{du^2}{\mathcal{F}}+dx^2+dy^2+\mathcal{H}dz^2\right)\,,\nn
\chi &= az, \qquad \phi = \phi(u), 
\end{align}
with trivial $X$-field, $X=1$ (i.e. $\psi=0$ in \eqref{defpsi}). The functions $\mathcal{F}, \mathcal{B}$ are functions of the radial coordinate $u$, and the function $\mathcal{H}$ is taken to be
$\mathcal{H} =  e^{-\phi}$ which, remarkably, can be imposed consistent with the equations of motion.

The black hole solutions constructed numerically in \cite{Mateos:2011ix,Mateos:2011tv} 
approach in the UV, located at $u\to 0$, a unit radius $AdS_5$ with a linear axion deformation.
As $T\to 0$ the black hole solutions
approach a $T=0$ domain wall solution whose IR limit
approaches a fixed point solution with
\begin{align}\label{liflike}
\mathcal{F} = \frac{49}{36}\left(\frac{12}{11}\right)^{\frac{6}{5}}{u}^{\frac{2}{7}},\qquad
\mathcal{B} = \mathcal{F}^{-1},\qquad
e^\phi = \left(\frac{11}{12}\right)^{\frac{2}{5}}{u}^{-\frac{4}{7}}\,,
\end{align}
which was first found in \cite{Azeyanagi:2009pr}.
After switching to a new radial coordinate $u=r^{-7/6}$, this solution 
can be written in the form
\begin{align}\label{aisotsol}
ds^2&=L^2\left(\frac{dr^2}{r^2}+r^2(-d\bar t^2+d\bar x^2+d\bar y^2)+r^{4/3}d\bar z^2\right)\,,\nn
\chi&=\bar a\bar z, \qquad e^{\phi}=L^{4/5}r^{2/3},
\end{align}
with $X=1$, where the bars denote quantities that have been rescaled, and $L^2=11/12$. This metric is manifestly invariant under the anisotropic Lifshitz-like
scaling $(\bar t,\bar x,\bar y,\bar z,r)\to (\lambda \bar t, \lambda \bar x,\lambda \bar y, \lambda^{2/3} \bar z,\lambda^{-1} r)$.

Following \cite{Azeyanagi:2009pr} we can study the properties of a massive scalar field, satisfying $\nabla^2\sigma=m^2\sigma$, in the background
\eqref{aisotsol}.
By considering solutions of the form $r^{\Delta_\pm}$ and demanding that $\Delta_\pm$ are real, we deduce that $m^2\ge -11/3$. 
Since we would like to
identify $\Delta_\pm$ as scaling dimensions in a putative field theory dual to these Lifshitz-like solutions, this suggests that the anisotropic solution \eqref{aisotsol} will be unstable under perturbations by any massive field with $m^2< -11/3$.
If the Lifshitz solution is unstable we expect that the $T=0$ domain
wall solution itself will be unstable and hence, by continuity, that the finite temperature black hole solutions will be unstable up to some critical temperature $T_c$.

Thus, since the scalar field $X$ in our model has $m^2=-4$, we anticipate that the 
black hole solutions of \cite{Mateos:2011ix,Mateos:2011tv} will still describe the high temperature phase of the system but will become unstable at $T_c$ leading to a phase transition. The critical temperature can be found by 
establishing the existence of a suitable zero-mode in the linearised
fluctuations of the $X$-field about the numerically constructed black holes of \cite{Mateos:2011ix,Mateos:2011tv}.
We carried out this analysis but we will omit the details. Instead we will focus on the construction of the new branch of fully
back reacted black holes that emerge at $T=T_c$ and examine some of the physical properties of the new low-temperature phase.

\section{Construction of new anisotropic black holes}\label{back-reacted}
\subsection{Ansatz and equations of motion}
We extend the ansatz of \cite{Azeyanagi:2009pr,Mateos:2011ix,Mateos:2011tv} 
by allowing for a non-trivial $X$-field and consider
\begin{align}\label{ansatz1}
ds^2 &= \frac{e^{-\frac{1}{2}\phi}}{u^2}\left(-\mathcal{FB}dt^2+dx^2+dy^2+\mathcal{H}dz^2+\frac{du^2}{\mathcal{F}}\right)\,,\nn
\chi &= az, \qquad \phi = \phi(u), \qquad X = {X}(u),
\end{align}
where 
$\mathcal{F}$, $\mathcal{B}$ and  $\mathcal{H}$ are functions of $u$. The function $\mathcal{H}$ is associated with the anisotropy in the $z$ direction 
that is sourced by the axion field. By combining the equation of motion for the dilaton with the Einstein equations, as in \cite{Azeyanagi:2009pr,Mateos:2011ix,Mateos:2011tv}, we find that it is possible to choose
the function $\mathcal{H}$ to be related to the dilaton via
\begin{equation}
\mathcal{H} =  e^{-\phi}\,,
\end{equation}
and we will do so in the sequel\footnote{Our preliminary investigations into relaxing this condition did not reveal any other solutions of physical
interest.}.

We now discuss the resulting equations of motion for this ansatz, following the approach of
\cite{Mateos:2011tv}. The equation for the axion $\chi$ in \eqref{5dEEquation} is trivially satisfied.
The $X$ equation of motion implies that
\begin{align}\label{xeom}
&12 u^2 \mathcal{F}  X{X}''   
+3\Big(-5 u^2 {X} \mathcal{F}  \phi ' 
+2 u^2 {X} \mathcal{F} \frac{\mathcal{B}'}{\mathcal{B}}
+4 u^2 {X}\mathcal{F}' -12 u {X}  \mathcal{F}    \Big){X}'\nn
&
-12 u^2  \mathcal{F}  \left({X}'\right)^2
+16 e^{-\phi /2}{X} (X^3-1) =0\,,
\end{align}
while the $\phi$ equation of motion gives
\begin{align}\label{phieom}
4 u \mathcal{F} \phi ''+2 u \mathcal{F}\frac{ \mathcal{B}' }{\mathcal{B}}\phi '+4 u \mathcal{F}' \phi '-5 u \mathcal{F} \left(\phi '\right)^2-12  \mathcal{F} \phi ' -4 a^2 u e^{3 \phi }= 0\,.
\end{align}
There are also four independent components of the Einstein equations arising from \eqref{5dEEquation}.
By taking a suitable combination of one of these equations with the $\phi$ equation of motion, in order to eliminate $\mathcal{B}' $ terms, we can arrive at
an equation which can be algebraically solved for $\mathcal{F}$:
\begin{align}\label{fsoleom}
\mathcal{F} = \frac{e^{-\frac{1}{2}{\phi}}}{4({\phi}'+u{\phi}'')}\left(e^{\frac{7}{2}{\phi}}a^2(4u+u^2{\phi}')+16{\phi}'\right)
+\frac{4e^{-\frac{1}{2}{\phi}}(1-{X})^2(2+{X})\phi'}{3{X}({\phi}'+u{\phi}'')}\,.
\end{align}
Next, by taking a suitable combination of two of the remaining three Einstein equations, in order to eliminate $\mathcal{F}''$ terms, we arrive at an equation that we can solve for ${\mathcal{B}'}/{\mathcal{B}}$:
\begin{align}\label{bpbeom}
\frac{\mathcal{B}'}{\mathcal{B}} = \frac{\left(24{\phi}'-9u{\phi}'^2+20u{\phi}''\right)}{24+10u{\phi}'}
-\frac{24 u {X}'^2}{{X}^2 \left(12+5u{\phi}'\right)}\,,
\end{align}
and we observe that only the combination ${\mathcal{B}'}/{\mathcal{B}}$ appears in \eqref{xeom} and \eqref{phieom}.
It is now possible to show that \eqref{xeom}, \eqref{phieom}, \eqref{fsoleom} and \eqref{bpbeom} imply that all of the Einstein equations are solved.
To see this we can use \eqref{phieom} to solve for $\mathcal{F}'$ in terms of $\phi$ and $X$ and their derivatives, after
using \eqref{fsoleom} and \eqref{bpbeom}. Furthermore, comparing this equation with the expression for $\mathcal{F}'$ that can be obtained by differentiating
 \eqref{fsoleom}, we obtain a third order equation for $\phi$, which can be used instead of \eqref{phieom}. One can
 then check that the remaining Einstein equations are satisfied. Observe that if we set $X=1$ we recover the equations of motion given in \cite{Mateos:2011tv}.

In summary, the equations of motion are equivalent to \eqref{xeom}-\eqref{bpbeom}
and are, effectively, second order in ${X}$, third order in $\phi$ and first order in $\mathcal{B}$, with 
$\mathcal{F}$ algebraically specified by $\phi$, 
${X}$ and their first and second derivatives. Thus, a solution is specified by six integration constants.

We note that the ansatz and hence the equations of motion are invariant under the following two scaling symmetries
\begin{align}\label{scsym}
&u\to \lambda u,\quad (t,x,y,z)\to \lambda(t,x,y,z),\quad a\to \lambda^{-1}a;\nn
&t\to \lambda t,\quad {\mathcal B}\to \lambda^{-1/2}{\mathcal B};
\end{align}
where $\lambda$ is a constant.

\subsection{The UV and IR expansions} 
We now discuss the boundary conditions that we will impose on \eqref{xeom}-\eqref{bpbeom}.
In the UV, as $u\to 0$, we demand that the asymptotic behaviour is given by
\begin{align}\label{uvexp}
{\phi} &= -\frac{a^2 u^2}{4}+u^4\frac{ \left(121 a^4+1152 \mathcal{B}_4+2304 (X_2)^2\right)}{4032}-u^4\log u\frac{a^4}{6}+\dots\,,\nn
\mathcal{F} &= 1+\frac{11 a^2 u^2}{24}+u^4\mathcal{F}_4 +u^4\log u \frac{7a^4}{12}  +\dots\,,\nn
\mathcal{B}&= 1-\frac{11 a^2 u^2}{24}+u^4\mathcal{B}_4-u^4\log u  \frac{7a^4}{12} +\dots\,,\nn
{X} &= 1+u^2 X_2-u^4\frac{5a^2X_2}{24}+\dots\,.
\end{align}
The solutions are asymptotically approaching $AdS_5$ with an anisotropic deformation of the axion field in the $z$-direction with strength $a$.
This UV expansion is specified by four parameters, ${\mathcal F_4}, {\mathcal B}_4, X_2$, whose physical interpretation will be discussed below,
and $a$. 
It is important to observe that we have set a possible $u^2\log u$ term in the expansion of $X$ to zero, as this would correspond to sourcing the operator dual to $X$
which
we don't want\footnote{In section \ref{critexp}, when we calculate critical exponents of the phase transition, we will 
briefly consider black holes with such a source for $X$.}. We next note that the second scaling symmetry in \eqref{scsym} has been used to set the leading term in $\mathcal{B}$ to unity.
We also observe that associated with the first scaling symmetry in \eqref{scsym} the UV expansion is preserved under
the transformations $u\to \lambda u$ and
\begin{align}\label{anomsc}
\mathcal{B}_4&\to \lambda^{-4}\mathcal{B}_4 +\frac{7}{12}a^4\lambda^{-4} \log \lambda\,, \nn
\mathcal{F}_4&\to \lambda^{-4}\mathcal{F}_4 -\frac{7}{12}a^4\lambda^{-4} \log \lambda\,, \nn
{X}_2 &\to\lambda^{-2}{X}_2\,.
\end{align}
The presence of the log terms is associated with the fact that the linear axion deformation gives rise to a non-vanishing conformal anomaly,
as discussed in \cite{Mateos:2011ix,Mateos:2011tv}.

In the IR, we will assume that we have a regular black hole horizon located at $u=u_h$. We therefore will demand that
as $u\to u_h$ we have
%\footnotesize
\begin{align}\label{irexp}
\phi &= \phi_h-\frac{ 12u_h X_h a^2e^{\frac{7 {\phi }_h}{2}}}
{32+3a^2e^{\frac{7 {\phi }_h}{2}}u_h^2X_h+16X_h^3}\left(u-u_h\right)+\dots\,,\nn
   \mathcal{F} &= {\mathcal F}_h\left(u-u_h\right) +\dots\,,\nn
\mathcal{B} &= \mathcal{B}_h+\frac{2 \mathcal{B}_h(45a^4e^{7\phi_h}u_h^4X_h^2-256(X_h^3-1)^2-96a^2e^{\frac{7\phi_h}{2}}u_h^2X_h(2+X_h^3))}{u_h\left(32+3a^2e^{\frac{7 {\phi }_h}{2}}u_h^2X_h+16X_h^3\right)^2}\left(u-u_h\right)+\dots\,,\nn
{X} &=X_h+\frac{16  X_h (X_h^3-1)}{u_h\left(32+3a^2e^{\frac{7 {\phi }_h}{2}}u_h^2X_h+16X_h^3\right)}\left(u-u_h\right)+\dots\,.
\end{align}
%\normalsize
This IR expansion is specified by four parameters, $\phi_h,{\mathcal B}_h,X_h$ and $u_h$, with ${\mathcal F}_h$ fixed via
\begin{align}
{\mathcal F}_h\equiv{\mathcal F}'(u_h)=-\frac{e^{-\frac{\phi_h}{2}}\left(32+3a^2e^{\frac{7 {\phi }_h}{2}}u_h^2X_h+16X_h^3\right)}{12u_hX_h}\,.
\end{align}

We have noted that the equations of motion are specified by six integration constants.
We have eight parameters appearing in the asymptotic expansion minus one for the remaining scaling symmetry 
in \eqref{scsym}. We thus expect to find a one-parameter family of solutions which can be parametrised by the quantity 
$T/a$. We note that the presence of the conformal anomaly introduces an additional dynamical scale which we
hold fixed to be unity throughout our discussion.

\subsection{Stress tensor and thermodynamics}

To calculate the free-energy and the stress tensor, we need to supplement the bulk action with boundary counter terms
(e.g. \cite{Skenderis:2002wp}). 
We write %in euclidean space
\begin{equation}
S_{total} = S_{bulk} + S_{ct}\,,
\end{equation}
where $S_{bulk}$ is the bulk action given in \eqref{5dAction} (or \eqref{5dAction2}) and, for the configurations of interest, we can take \cite{Bianchi:2001de,Papadimitriou:2011qb}
\begin{equation}
\begin{split}\label{ct}
S_{ct} = &\int\ \mathrm{d}^4x\sqrt{-\gamma}\left(2K - 6+\frac{1}{4}e^{2\phi}\partial_i\chi \partial^i\chi-\psi^2\left(1+\frac{1}{2 \log u}\right)\right)
+\log u\int\ \mathrm{d}^4x\sqrt{-\gamma}\mathcal{A}
% +\frac{1}{4}(c_{sch}-1)\int\ \mathrm{d}^4x\sqrt{\gamma}\mathcal{A}
\end{split}
\end{equation}
where $\partial^i = \gamma^{ij}\partial_j$ and
$\mathcal{A}$ is the conformal anomaly in the axion-dilaton-gravity system given by \cite{Papadimitriou:2011qb} 
\begin{align}\label{anom}
\mathcal{A}=\frac{1}{6}e^{4\phi}|\partial\chi|^4\,.
\end{align}
Note that here we have expressed the $X$ scalar field in terms of the canonically normalised scalar $\psi$ defined by
$X = e^{-\frac{\psi}{\sqrt{6}}}$, which we will continue to use throughout this section. 
We note that the $1/\log u$ term is only relevant for solutions where the $X$-field is sourced, which are only 
briefly discussed in section \ref{critexp}.

The expectation value of the stress energy tensor is obtained by taking the functional derivative of the total action with
respect to the boundary metric \cite{Henningson:1998ey,Balasubramanian:1999re}
\begin{align}
T^{ij}&=\lim_{u\to 0}\Bigg(-2K^{ij}+\gamma^{ij}\left(2K -6+\frac{1}{4}e^{2\phi}\partial_i\chi \partial^i\chi-\psi^2(1+\frac{1}{2 \log u})\right)
\nn
&\qquad\qquad-\frac{1}{2}e^{2\phi}\partial^i\chi\partial^j\chi+\log u\left( \mathcal{A}\gamma^{ij}-\frac{2}{3} e^{4\phi}\partial^i\chi\partial^j\chi (\partial\chi)^2\right)\Bigg)\,,
\end{align}
where $K_{ij}$ is the extrinsic curvature of a $u =$ constant hypersurface, and $\partial^i = \gamma^{ij}\partial_j$.
Using the boundary expansion of the fields in the previous section, we find the expectation value of the stress-energy tensor 
has the following non-vanishing components:
\begin{align}\label{stress}
T^{tt} &= \left(-3\mathcal{F}_4-\frac{23}{7}\mathcal{B}_4+\frac{2945}{4032}a^4 -\frac{4}{7}(X_2)^2\right)\,,\nn
T^{xx}=T^{yy} &= \left(-\mathcal{F}_4-\frac{5}{7}\mathcal{B}_4+\frac{443}{4032}a^4+ \frac{4}{7}(X_2)^2\right)\,,\nn
T^{zz}&= \left(-\mathcal{F}_4-\frac{13}{7}\mathcal{B}_4+\frac{2731}{4032}a^4  -\frac{12}{7}(X_2)^2\right)\,.
\end{align}
This result is consistent with \cite{Mateos:2011tv} when $X_2\rightarrow 0$.

Similarly, we can calculate the one-point functions of the theory in order to find the vacuum expectation of the fields. For the scalar fields
we find that expectation values and sources are given by 
\begin{align}
\langle\mathcal{O}_\chi\rangle &= 0,  &\chi_{(0)} =az\,,\nn
\langle\mathcal{O}_\phi\rangle &=- \frac{143}{252}a^4+\frac{8}{7}(\mathcal{B}_4+2X_2^2), &\phi_{(0)} = 0\,,\nn
%\langle\mathcal{O}_X\rangle \equiv
\langle\mathcal{O}_\psi\rangle &= -\sqrt{6}X_2,  &\psi_{(0)} = 0\,.
\end{align}

We can now easily check that the Ward identities for the theory are satisfied. 
Firstly, diffeomorphism invariance gives us the conservation of the stress-energy tensor
\begin{equation}
\nabla^i T_{ij} +\langle \mathcal{O}_\phi\rangle \nabla_j\phi_{(0)}  +\langle \mathcal{O}_\chi\rangle\nabla_j\chi_{(0)}  +\langle \mathcal{O}_\psi\rangle \nabla_j\psi_{(0)} =0\,,
\end{equation}
which in our case is simply $\nabla^iT_{ij}=0$, and is trivially satisfied.
Similarly, the invariance of the theory under Weyl transformations leads to the conformal Ward anomaly
\begin{equation}
T^i_i =-(4-\Delta_\psi)\psi_{(0)}\langle\mathcal{O}_\psi\rangle+\mathcal{A}\,,
\end{equation}
where $\mathcal{A}$ is the conformal anomaly. From \eqref{stress} we have $T^i_i =a^4/6$ and hence
\begin{equation}
\mathcal{A} = \frac{a^4}{6}\,,
\end{equation}
in agreement with a direct calculation of \eqref{anom}.

By analytically continuing the time coordinate via $t=-i\tau$ and demanding regularity of the metric at $u=u_h$, 
we find that
he Hawking temperature of the black holes is given by
\begin{align}\label{htemp}
T=\frac{{\mathcal B}_h^{1/2}|{\mathcal F}_h|}{4\pi}\,.
\end{align}
The entropy density of the black holes, $s$, can be obtained from the area of the 
black hole horizon and since we have set $16\pi G=1$ we have
\begin{equation}\label{ent}
s = 4\pi\frac{e^{-\frac{5}{4}\phi_h}}{u_h^3}.
\end{equation}
We can calculate the free-energy density, $w$, by calculating the total on-shell Euclidean action via $w\text{vol}_3=TI_{total}|_{os}$. 
In fact using the results of \cite{Donos:2013cka} we can immediately obtain 
\begin{align}\label{sm1}
w=E-Ts\,,
\end{align}
where $E=T^{tt}$ as well as the Smarr formula
\begin{align}\label{smarr}
E-Ts=-T^{xx}\,.
\end{align}
As we explain in the appendix, these results can also be obtained by explicitly writing the bulk action as a total derivative in two different ways.

Finally, we note that we can determine how various quantities transform under the scaling given in \eqref{anomsc}. For example, the free-energy transforms
as
\begin{align}
w\to \lambda^{-4}w-{\mathcal A}\lambda^{-4}\log\lambda\,.
\end{align}
One can check that the Smarr formula is invariant under \eqref{anomsc}.

\subsection{Numerical construction of the black hole solutions}
We construct the black hole solutions by numerically solving the ODEs \eqref{xeom}-\eqref{bpbeom},
subject to the boundary conditions given in \eqref{uvexp},\eqref{irexp}. Recall that for a fixed dynamical scale the black hole solutions can be parametrised by
$T/a$. In practise we set $a=1$ and use a numerical shooting method in which we shoot from both near the black hole horizon and the holographic boundary
and then match in the middle.

\begin{figure}
\centering
\includegraphics[scale = 0.6]{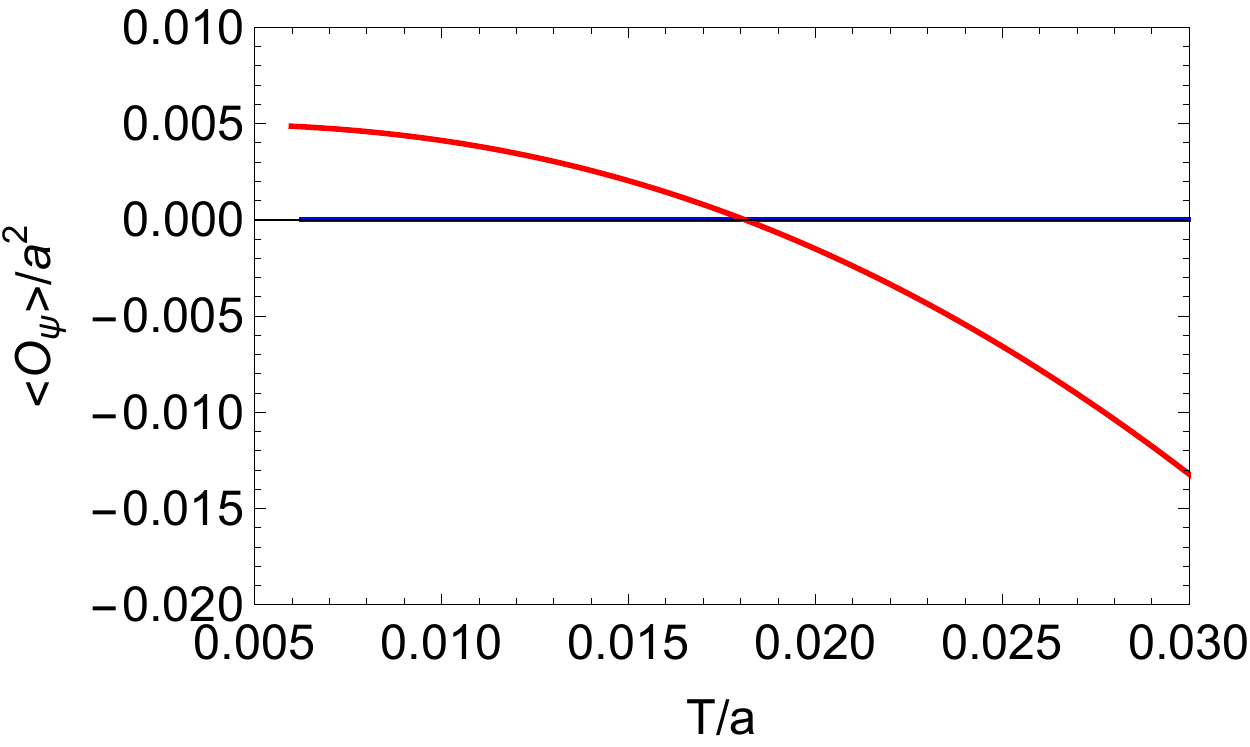} 
\caption{Plot showing the expectation value of the operator ${\cal O}_\psi$ dual to
the scalar field $\psi$ (recall $X=e^{-{\psi}/{\sqrt 6}}$)
for the black hole solutions.
The blue line corresponds to the solution of \cite{Mateos:2011tv}, while the red lines indicate the two new branches of solution, one
that exists for $T\le T_c$ and the other which appears to exist only for $T\ge T_c$.
}
\label{XSolution}
\end{figure}

\begin{figure}
\centering
\includegraphics[scale = 0.58]{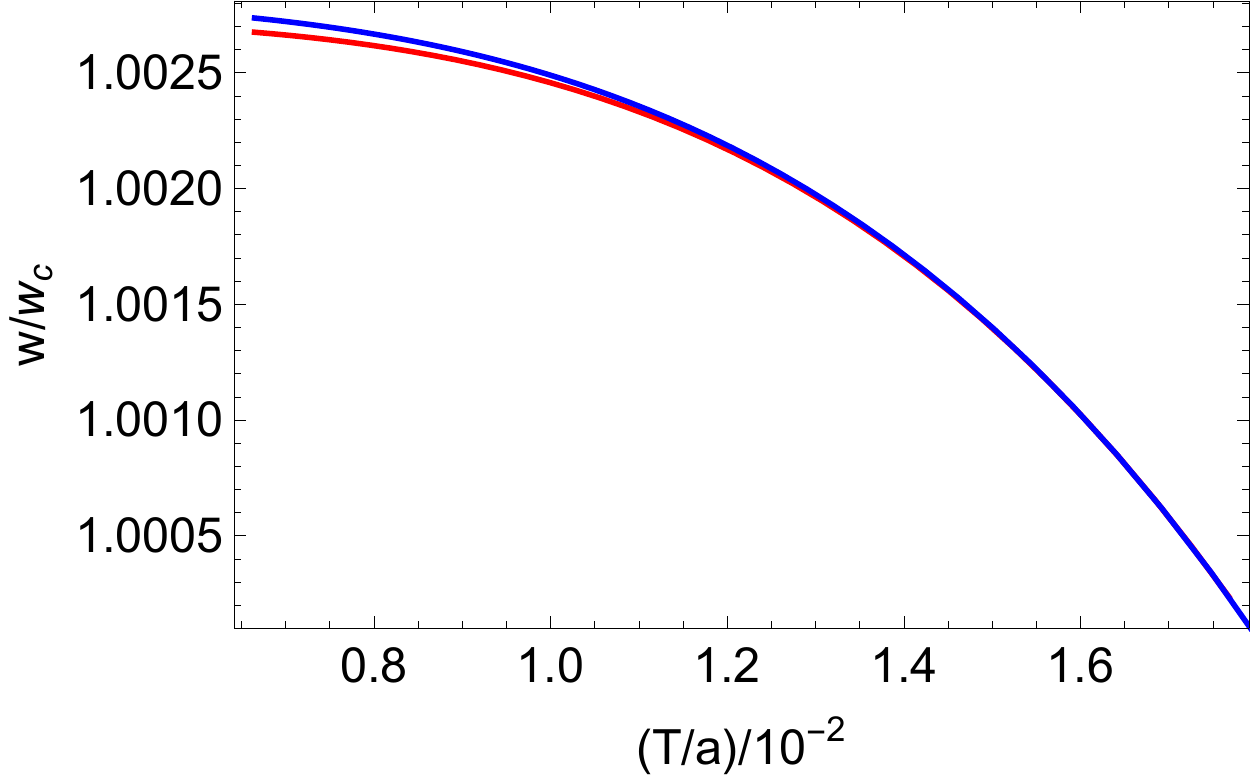}  \,
\includegraphics[scale = 0.56]{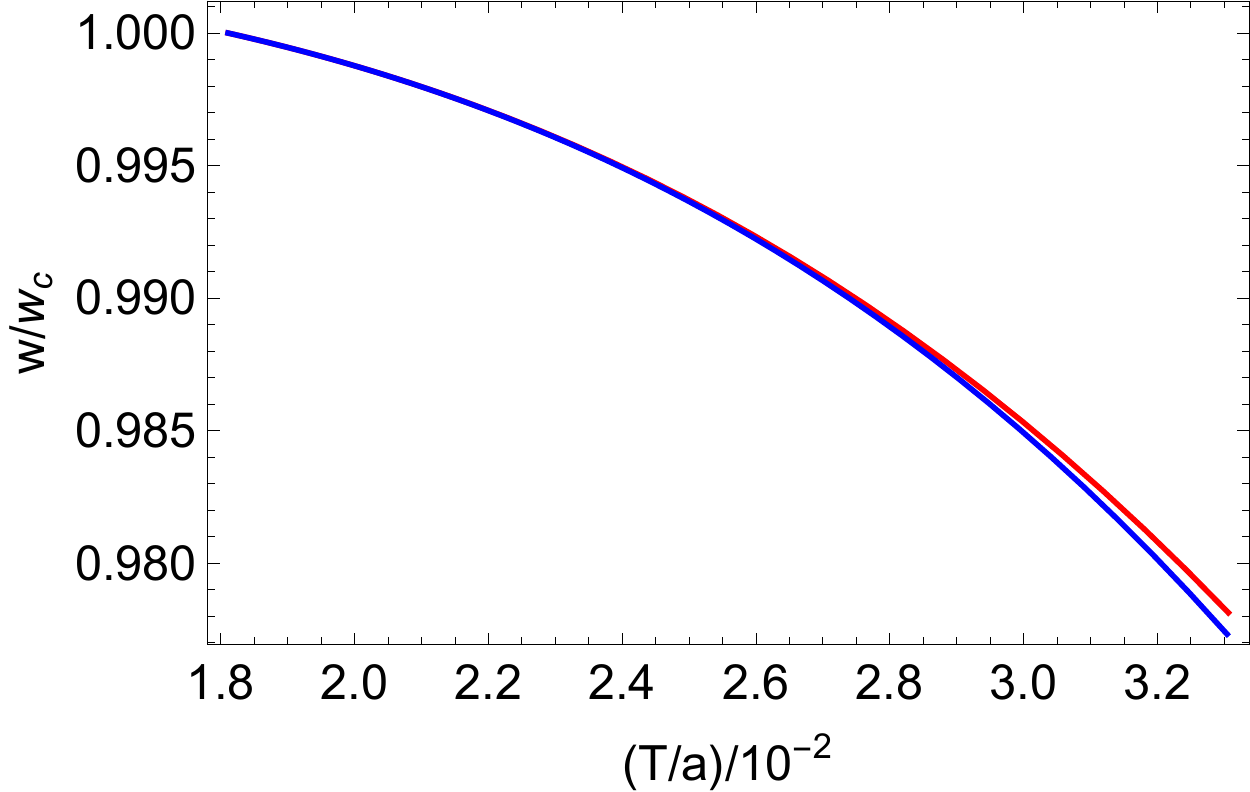} 
\caption{Plot showing the free energy of the black hole solutions, 
relative to the free energy at the critical temperature, $w_c$. The red line is the new branch of solution, while the blue line is the solution of \cite{Mateos:2011tv}. 
The left panel shows that the of branch of black holes that exist for $T\le T_c$ has lower free energy, and hence is thermodynamically preferred. The right panel shows the black holes with $T\ge T_c$ are not preferred. }\label{FEPlot}
\end{figure}

We find that a new branch of black hole solutions appears at the critical temperature $T_c/a\sim 1.8\times 10^{-2}$.
In fact we find that there exist two distinct branches of black hole solutions carrying $X$ hair, 
one that exists for $T\le T_c$ and the other that exists for $T\ge T_c$, as illustrated in figures \ref{XSolution} and \ref{FEPlot}.

The solutions with $T\le T_c$ are the physically relevant solutions. In particular, we see from figure \ref{FEPlot}
that for $T\le T_c$, where the black hole solutions of \cite{Mateos:2011ix,Mateos:2011tv} are unstable, this new branch of solutions
has lower free energy. We thus conclude that there is a phase transition which, moreover, is a continuous phase transition. We emphasise that these hairy black holes
are associated with a spontaneous phase transition since the boundary conditions we imposed for the field $X$ corresponded to the dual
operator ${\cal O}_\psi$ acquiring an expectation value with no source. From the point of view of the $D=5$ model this new phase does not appear to break any more symmetries than
the background black holes. In particular, one can see from that potential in \eqref{5dAction2} does not have, for example, a $\mathbb{Z}_2$ symmetry. However, after uplifting to type IIB, following an earlier discussion we know that when the $\psi$-field acquires an expectation value then the $SO(6)$ global symmetry is spontaneously broken to $SO(4)\times SO(2)$.

The black hole solutions with $T\ge T_c$ appear to be ``exotic hairy black holes". In particular, they only seem\footnote{We have checked that this is true up to $T/a\sim 3$.} 
to exist for
$T\ge T_c$, in contrast to black holes associated with a first order transition which start existing for $T\ge T_c$ and then turn around at some maximum temperature before continuing down to lower temperatures. We also observe from figure \ref{FEPlot} that these black holes have higher free energy than the black holes of \cite{Mateos:2011ix,Mateos:2011tv} and hence are not thermodynamically preferred.
We note that such exotic hairy black holes have appeared in other holographic constructions, both bottom up \cite{Buchel:2009ge} 
and top-down \cite{Donos:2011ut}.

\subsection{Critical Exponents}\label{critexp}

Having shown that there is a continuous phase transition at $T_c$, we now investigate the critical exponents of the 
transition. Somewhat surprisingly, we find that the phase transition does not
have the same critical exponents as the majority
of holographic phase transitions.

The simplest critical exponent to calculate is $\beta$, which is defined by
\begin{equation}
\langle \mathcal{O}_\psi \rangle \sim (T_c - T)^\beta \,.
\end{equation}
For our phase transition, from our numerics we find that $\beta=1$, differing from the standard value $\beta =1/2$.
There are several other important critical exponents for a phase transition\footnote{For a discussion in the context of holography see \cite{Maeda:2009wv}.}. For example, the 
behaviour of the specific heat for $T<T_c$ defines the
exponent $\alpha$ via
\begin{align}
C \sim (T_c-T)^{-\alpha}\,.
\end{align}
This can be read off from the behaviour of the difference between the free energies
of the two phases via $\Delta w\sim (T_c-T)^{2-\alpha}$. We find that in our transition $\alpha=-1$ in contrast to the standard value of $\alpha=0$. The remaining critical exponents are fixed by $\alpha,\beta$ using scaling relations. For example
we have
\begin{align}
\gamma=2-\alpha-2\beta^{-1},\qquad \delta=(2-\alpha)\beta^{-1}-1\,,
\end{align}
where $\gamma,\delta$ are defined by
\begin{align}
\frac{\partial \langle \mathcal{O}_\psi \rangle}{\partial \psi} \sim (T_c-T)^\gamma,\qquad
\psi \sim \langle {\mathcal O}_\psi \rangle^\delta
\end{align}
For our black holes we obtain $\gamma=1,\delta=2$
 in contrast to the standard results of $\gamma=1,\delta=3$.

As a check that the scaling relations are indeed satisfied, we carried out a direct calculation of the exponent $\delta$. To do this we constructed a more general class of
black hole solutions with a source for the operator dual to $\psi$. This required changing the boundary conditions
in \eqref{uvexp} to allow for terms of the form $u^2\log u$. Having done this (which requires some effort),
it is possible to see how the behaviour
of $\langle {\mathcal O }_\psi \rangle$ needs to be changed as one switches on the source, while keeping the temperature fixed to be at  the value $T=T_c$.
Carrying out this procedure we found $\delta=2$ in agreement with above.
To summarise, the critical exponents for the new phase of black holes are
\begin{align}\label{newexps}
(\alpha,\beta,\gamma,\delta)=(-1,1,1,2)\,,
\end{align}
in contrast to the standard values $(\alpha,\beta,\gamma,\delta)=(0,1/2,1,3)$.
Recall that the standard values arise from a Landau-Ginzburg model with quadratic and quartic terms in the free energy. In appendix B we discuss how the exponents
\eqref{newexps} are associated with a free energy for a scalar order parameter with quadratic and cubic terms.

Some bottom-up holographic superconducting phase transitions have been studied in the probe 
approximation \cite{Franco:2009yz,Franco:2009if,Herzog:2010vz} and with back-reaction \cite{Aprile:2010yb},
which also exhibit
non-standard critical exponents. The specific critical exponents that we have found for our new black holes have
also been found in models
with a potential which contained terms that are cubic in the modulus of a complex scalar field. 
Although such couplings are rather unnatural for a charged scalar field, and it is difficult to see how they would arise form a top-down setting,
we find that cubic terms in the potential for the neutral scalar field $\psi$ in our model are responsible for the non-mean field behaviour.

To see this we first note that if we expand our Lagrangian for the scalar field $\psi$ around $\psi=0$ we have
\begin{align}
\label{fullexp}
\mathcal{L}_{\psi} 
&= -\frac{1}{2}(\partial\psi)^2 + 12 + 2 \psi^2 -\frac{1}{2}\sqrt{\frac{2}{3}}\psi^3 + \frac{1}{12}\psi^4 + \mathcal{O}(\psi^5) , 
\end{align}
In particular, the absence of a $\mathbb{Z}_2$ symmetry $\psi \rightarrow - \psi$ allows for the cubic term.
We can contrast this model with a bottom up model in which the cubic term is absent:
\begin{equation}\label{meanL}
\mathcal{L}_{\psi} = -\frac{1}{2}(\partial\psi)^2 + 12 + 2 \psi^2 + \frac{1}{12}\psi^4 + \mathcal{O}(\psi^5) , 
\end{equation}
with the rest of the Lagrangian unchanged. 
We have constructed the back-reacted black holes for this model and we find that the
critical exponents for the phase transition now take the standard values.
Furthermore, we have checked that varying the quartic terms does not change this result.

\subsection{Zero temperature scaling solution}\label{zerotemp}
\begin{figure}
\includegraphics[scale = 0.625]{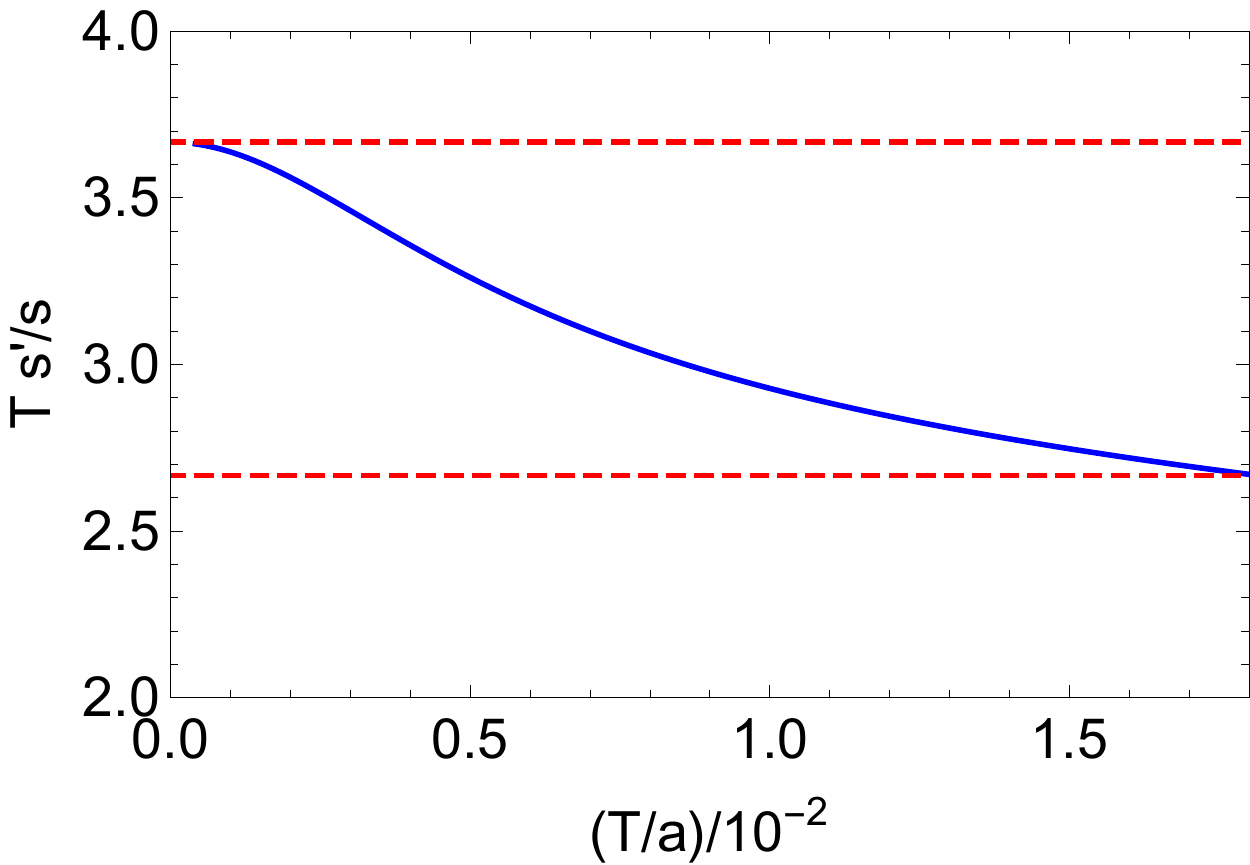} \,
\includegraphics[scale = 0.65]{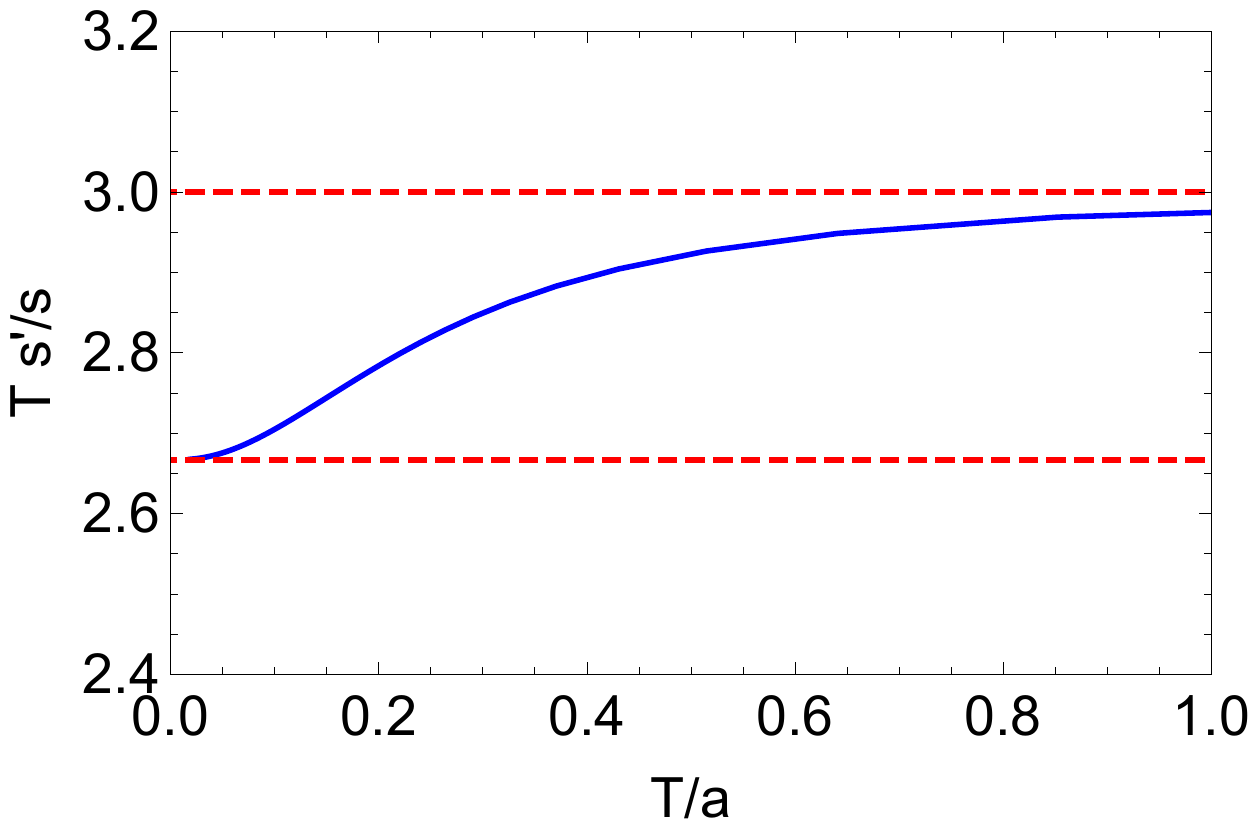} 
\caption{Plots showing the temperature dependence of entropy, both along the high temperature branch of solution (right), and for the new low temperature branch of solution (left). The dotted lines are $11/3$ and $8/3$ on the left plot, and $3$ and $8/3$ for the right plot. We see that at very high temperatures $s$ exhibits the standard scaling of 
the $N=4$ SYM plasma, before cooling towards a Lifshitz geometry. After the phase transition, the system moves away from the Lifshitz geometry, and towards the new scaling solution.}\label{EntPlot}
\end{figure}

To investigate the low temperature behaviour of black hole solutions
it is often illuminating to
examine the low temperature behaviour of $Ts'/s$, since if it approaches a constant it indicates an emergent scaling behaviour which one
can then try to identify. 
For the black hole solutions constructed in \cite{Mateos:2011tv} with $X=0$ one finds that 
$s$ scales as $s \sim T^{\frac{8}{3}}$ and this is exactly the scaling behaviour that is associated with the 
Lifshitz-like anisotropic geometry found by \cite{Azeyanagi:2009pr}
that appears at $T=0$ in the far IR. In fact this scaling behaviour is approximately present at the critical temperature
phase transition as we see from figure \ref{EntPlot}.

For our new black hole solutions with $X\ne 0$ we also see from figure \ref{EntPlot} that at very low temperatures $s \sim T^{\frac{11}{3}}$. 
We therefore look for the existence of a scaling solution to the equations of motion of the form
\begin{gather}
e^{\phi(u)} = e^{\phi_0}u^{\phi_c}, \, \, \,  \mathcal{F}(u) = \mathcal{F}_0u^{\mathcal{F}_c}, \,\,\, \mathcal{B}(u) = \mathcal{B}_0u^{\mathcal{B}_c}, \,\,\, X(u) = X_0u^{X_c} \,.
\end{gather}
However, by analysing the resulting algebraic equations one concludes that when $X\ne 0$ such solutions do not exist.
Instead, we have found that the equations of motion admit the following expansion as $u \rightarrow \infty$:
\begin{align}\label{scalsol}
e^{\phi(u)} &=  \frac{\phi_0}{ (au)^{4/9}} -  \frac{15232}{10935  (au)^{16/9}\phi_0^{19/2}} + \dots\,,\nn
\mathcal{F}(u) &= \frac{81}{112}\phi_0^3(au)^{2/3} + \frac{28}{45\phi_0^{15/2}(au)^{2/3}} + \dots\,,\nn
 \mathcal{B}(u) &= \frac{\mathcal{B}_0}{(au)^{2/3}} -\frac{3136\mathcal{B}_0}{3645 (au)^2 \phi_0^{21/2}}+\dots\,,\nn
 X(u) &= \frac{4}{3(au)^{4/9}\phi_0^{7/2}} - \frac{28672}{32805  (au)^{16/9} \phi_0^{14} } + \dots\,,
\end{align}
where
 $\mathcal{B}_0, \phi_0$ are constant.
Furthermore, we have checked that the new black hole solutions start to approach this behaviour at low temperatures.
Moreover, we can also show that this behaviour is associated with the observed scaling,
$s \sim T^{\frac{11}{3}}$.

To see this we first observe that the above expansion can be generalised to finite temperatures, with the leading order expansion of $\mathcal{F}$ replaced with 
\begin{equation}
\mathcal{F}(u) = \frac{81}{112}\phi_0^3 (au)^{2/3}\left(1-\left(\frac{u}{u_h}\right)^{28/9}\right) + ...\, ,
\end{equation}
where $u_h$ is the horizon radius. By combining (\ref{ent}) and the above finite temperature solution, the entropy is given by
\begin{equation}
s = {4\pi}a^3\phi_0^{-5/4} (au_h)^{-22/9} + ...
\end{equation}
while the Hawking temperature is given by
\begin{equation}
T =\frac{9a\sqrt{\mathcal{B}_0}\phi_0^3}{16\pi}(au_h)^{-2/3} + ... \,\, .
\end{equation}
Combining these two expressions we find, as claimed:
\begin{equation}
s = \left(\frac{4}{3}\right)^{1/3}\frac{65536  \pi^{14/3}a^3}{ 2187 \phi_0^{49/4}\mathcal{B}_0^{11/6}}\left(\frac{T}{a}\right)^{11/3}+ ... \,.
\end{equation}

It is interesting to point out that the leading behaviour of the $T=0$ solution given
in \eqref{scalsol} can be recast in the following form, after making the coordinate transformation
$u=c\rho^{3/2}$, for some constant $c$:
\begin{align}
&ds^2\sim\rho^\frac{-2(3-\theta)}{3}\left(d\rho^2-d\bar t^2+d\bar x^2+d\bar y^2+\rho^{-2(z-1)}d\bar z^2\right)\,,\nn
&e^\phi\sim\rho^{-2/3},\qquad X\sim \rho^{-2/3}\,,
\end{align}
with $\theta=-1$, $z=2/3$ and the bars denote that we have rescaled the coordinates. This is similar to the
hyper-scaling solutions with Lifshitz exponent $z$ and a hyper-scaling violation exponent $\theta$ \cite{Charmousis:2010zz,Ogawa:2011bz,Huijse:2011ef}, but here the Lifshitz exponent is associated with a spatial direction and not a time direction.
Under the scaling 
$(\bar t,\bar x,\bar y,\bar z,\rho)\to (\lambda \bar t, \lambda \bar x,\lambda \bar y, \lambda^{2/3} \bar z,\lambda\rho)$ we find that
metric transforms as $ds\to \lambda^\frac{\theta}{3}ds$. It is curious that the exponent $z=2/3$ is the same value as
for the unstable Lifshitz-like ground state with $X=1$ given in \eqref{aisotsol}.

\subsection{Thermal conductivity}
Having established the low temperature behaviour of the black holes, it is of interest to derive the DC thermal conductivity 
in the $z$-direction\footnote{Since the solutions are still translationally invariant in the $x$ and $y$ directions, the thermal conductivity in these directions  is infinite.}, $\kappa$. To do so we follow the approach of \cite{Donos:2014cya} which showed how $\kappa$ can be obtained in terms of black hole
horizon data. 

To make contact with \cite{Donos:2014cya} it is convenient to write the black hole solutions in a slightly different form
\begin{align}
ds^2 &= -Udt^2 + \frac{dr^2}{U} + e^{V_1}(dx^2+dy^2) + e^{V_3}dz^2\,,\nn
\chi&=a z, \qquad \phi = \phi(r), \qquad X = X(r)
\end{align}
where $U,V_1$ and $V_3$ are functions of $r$. We assume that as $r \rightarrow \infty$, the functions have the following asymptotic form
\begin{gather}
U \sim r^2 + ..., \quad e^{2V_i} \sim r^2 + ... ,
\end{gather}
and $\phi\to 0+o(u^2)$, $X\to 1+o(u^2)$.
We now consider a small linearised perturbation about this class of black hole solutions that includes a piece that is linear in time:
\begin{align}\label{dcxan2}
g_{tz}&=t\delta f_2(r)+\delta g_{tx_1}(r)\,,\nn
g_{rz}&=e^{{2V_e}}\delta h_{rz}(r)\,,\nn
\chi_1&=az+\delta\chi_1(r)\, .%\qquad \chi_2=kx_2+\delta\chi_2(t,r)\,,
\end{align}

A key point is that this perturbation does not source the $X$-field. As a result the calculation of $\kappa$ is virtually unchanged from
the derivation given in \cite{Donos:2014cya}. Rather than repeat the steps, we just quote the final result:
\begin{equation}
\kappa %= \frac{1}{T}\frac{\partial}{\partial\zeta}Q 
= \left[\frac{4\pi sT }{a^2e^{2\phi}}\right]_{r = r_h} \, .
\end{equation}

We showed in the previous section that the black holes with $X\ne 0$ have $s \sim T^{11/3}$ at low temperatures. From (\ref{scalsol}) we can also determine the low temperature scaling behaviour of the dilaton to be $(e^{2\phi})_{r = r_h} \sim T^{4/3}$. Hence the low temperature scaling of the thermal conductivity is given by $\kappa \sim T^{10/3}$. We see that the the black hole solution is dual to a ground state that is thermally insulating in the direction of the linear axion. It is also interesting to contrast this result with the result for the (unstable)
Lifshitz ground state, where $\kappa \sim T^{7/3}$\cite{Donos:2014cya}.

\section{Discussion} \label{concl}

We have shown that the anisotropically deformed $N=4$ Yang-Mills plasma studied in \cite{Mateos:2011tv,Mateos:2011ix} has low temperature instabilities. The plasma undergoes a third-order phase transition, spontaneously breaking the global $SO(6)$ symmetry down to $SO(4)\times SO(2)$. 
We showed that critical exponents of the phase transition are given by
$(\alpha,\beta,\gamma,\delta)=(-1,1,1,2)$ in contrast to the standard
mean field theory values usually seen in holography. These critical values can be associated with a cubic
Landau-Ginzburg free energy for a scalar order parameter, as discussed in appendix B. However, such
a free energy is unstable. In addition stabilising the free energy with a higher powers of the
order parameter leads to a first order phase transition. By contrast, in our holographic model
the transition appears to be continuous, in fact third order, provided that the
branch of black holes with $T>T_c$ does not turn around at some temperature
and then go down to lower energies. Thus, our model underscores the
difficulty in making a precise identification of the properties of the phase transition
just using a Landau-Ginzburg mean field approach. Perhaps a Landau-Ginzburg
model with more fields might give a better description.
It would be interesting to further clarify this point.

It would also be interesting to know which critical exponents can be realised in string/M-theory constructions. In addition to the exponents that we found here
it was shown that for a class of top-down $R$-charged black holes the critical exponents are given by $(\alpha,\beta,\gamma,\delta)=(1/2,1/2,1/2,2)$ \cite{Cai:1998ji,Cvetic:1999rb,Maeda:2008hn}. It is not clear if the bottom up 
constructions discussed in \cite{Franco:2009yz,Franco:2009if,Herzog:2010vz,Aprile:2010yb}, which had more general exponents, can be embedded into top-down setting.
Following  \cite{Maeda:2009wv}, it would also be of interest to explicitly calculate the dynamic critical exponents for our transition as well as others.

We analysed the $T=0$ limiting behaviour of the black holes describing the new low temperature phase. We showed that in the far IR there is an emergent leading order behaviour that is similar
to the hyperscaling violation geometries but with spatial anisotropic scaling. This scaling behaviour implies that the thermal conductivity scales with temperature as $\kappa\sim T^{10/3}$ at low temperatures, revealing that the ground state is a thermal insulator.

The black hole solutions were constructed using a consistent KK truncation that keeps a single scalar field $X$ with $m^2=-2$. This scalar field is
part of a multiplet of twenty scalars that transform in the ${\bf 20}'$ of $SO(6)$.  
All of these scalars become unstable at the critical temperature $T_c$ and it would be very interesting to investigate the full class of black hole solutions
that emerges at $T_c$, which will generically break all of the $SO(6)$ symmetry, 
and then follow them down to low temperatures. Although challenging, this could be investigated using the consistent truncation of
\cite{Cvetic:2000nc} that keeps twenty scalars parametrised by a symmetric, unimodular six by six matrix $T_{ij}$. 
As a first step one could analyse the truncation that keeps five scalar fields, parametrised by the diagonal subset \cite{Cvetic:2000eb}, or even simpler, the truncations that keeps 
just two scalar fields \cite{Cvetic:1999xp}. 
%We will leave this interesting topic for future work.
The spontaneous breakdown of the global $SO(6)$ symmetry will lead to
Goldstone modes and it would also be interesting to study these further, both from
the gravitational and field theory points of view.

Finally, it is not difficult to show that the $D=5$ model with metric, axion and dilaton (i.e. when $X=1$)
arises as a consistent truncation on an arbitrary five-dimensional Sasaki-Einstein (SE) manifold, not
just the five-sphere. Therefore, the original black hole solutions of \cite{Azeyanagi:2009pr,Mateos:2011ix,Mateos:2011tv} also describe the high temperature
phase of the whole class of dual $N=1$ SCFT plasmas with an anisotropic deformation. 
For a given SE space, if  there are no BF saturating modes in the spectrum
then the black holes will not suffer the instabilities that we have described in this paper, and the Lifshitz ground
state constructed in \cite{Azeyanagi:2009pr} may be the true ground state of the system. On the other hand if there are BF saturating modes then
the black holes will become unstable at some critical temperature. For a general SE manifold it is unlikely that there
is a consistent truncation maintaining just one extra scalar field as we have studied in this paper
for the case of the five-sphere. This would mean that the corresponding black hole solutions describing
the low temperature phase would need to be constructed directly in ten spacetime dimensions.
Although this is likely to be a challenging task, it may be tractable to study the solutions near the phase
transition and it would be particularly interesting to determine the critical exponents.

\section*{Acknowledgements}
We thank Aristomenis Donos, Sebastian Franco, Chris Herzog and especially
Makoto Natsuume
for helpful discussions.
The work is supported by STFC grant ST/J0003533/1, EPSRC programme grant EP/K034456/1,
and also by the European Research Council under the European Union's Seventh Framework Programme (FP7/2007-2013), ERC Grant agreement
ADG 339140.
\appendix
\renewcommand{\thesection}{\Alph{section}}

\section{Smarr relation}

We explain how to obtain the Smarr relation \eqref{smarr} via a direct calculation of the on-shell action.
The bulk Euclidean bulk action is given by
\begin{equation}
I_{bulk}=-\Delta\tau\mathrm{vol}_3\int^{u_h}_0\mathrm{d}u{\cal L}_{bulk}\,,
\end{equation}
where the Lagrangian density  integrand is given by
\begin{align}
{\cal L}_{bulk}=\sqrt{-g}\left(R-3X^{-2}(\partial X)^2+4(X^2+2X^{-1}) -\frac{1}{2}(\partial\phi)^2-\frac{1}{2}e^{2\phi}(\partial\chi)^2\right)\,.
\end{align}

We would like to rewrite this as a total derivative in $u$ after using the equations of motion. To achieve this we found it helpful to
use the fact that after contraction of equation (\ref{5dEEquation}), we can write the integrand of the action as
\begin{equation}
{\cal L}_{bulk}=-\frac{8 \sqrt{\mathcal{B}} e^{-\frac{7 \phi }{4}}}{3 u^5}\left(X^2+2X^{-1}\right)\,.
\end{equation}

After some work we find that after using the equations of motion
\eqref{xeom}-\eqref{bpbeom} the integrand can be written as
\begin{equation}
\begin{split}
{\cal L}_{bulk}=&\left(2\frac{\sqrt{\mathcal{B}} \mathcal{F} e^{-5 \phi /4}}{u^4}-\frac{\mathcal{F} e^{-5 \phi /4} \mathcal{B}'}{u^3 \sqrt{\mathcal{B}}}-\frac{\sqrt{\mathcal{B}} e^{-5 \phi /4} \mathcal{F}'}{u^3}+\frac{1}{2}\frac{\sqrt{\mathcal{B}} \mathcal{F} e^{-5 \phi /4} \phi '}{u^3}\right)'\,,
\end{split}
\end{equation}
where the prime indicates differentiation with respect to the $u$ coordinate. Using this expression we will get contributions to the on-shell action
both from the horizon and the boundary. Using the near horizon and boundary expansions of the fields given in \eqref{uvexp},\eqref{irexp}, 
and combining with the boundary counter terms we deduce that the free energy density can be expressed as
\begin{equation}
w =E - sT\,,
\end{equation}
as in \eqref{sm1}.

On the other hand, using \eqref{xeom}-\eqref{bpbeom} we can also write the integrand in the form
\begin{equation}
\begin{split}
{\cal L}_{bulk}=&\left(2\frac{\sqrt{\mathcal{B}} \mathcal{F} e^{-5 \phi /4}}{u^4}+\frac{\sqrt{\mathcal{B}} \mathcal{F} e^{-5 \phi /4} \phi '}{2u^3}\right)'\,.
\end{split}
\end{equation}
This only gives contributions from the boundary leading to
\begin{equation}
w = -T^{xx} \,.
\end{equation}
Combining these gives these expressions gives the Smarr relation \eqref{smarr}.

\section[Critical exponents for a cubic free energy]{Critical exponents for a cubic free energy\footnote{We would like to thank Makoto Natsuume for helpful discussions on this section.}}
Suppose we have a Landau-Ginzburg free energy functional for a scalar order parameter, $m$, of the form
\begin{align}\label{felg}
f=f_0+\frac{am^2}{2}+\frac{bm^3}{3}\,,
\end{align}
with $f_0$ a constant, $a=t^n$, with $t=(T-T_c)/T_c$, and $b$ is a temperature
dependent constant which we take to be positive. We choose $n$ so that $a<0$ for $T<T_c$ and we will be especially interested in the case $n=1$.
For the moment let us ignore the global instability for $m<0$ and focus on the extrema at $m=0$ and $m=-a/b$ which exists when $a<0$ i.e for $T<T_c$. For the latter minimum we have $m\propto t^n$ and hence we conclude that $\beta=n$. To obtain $\alpha$ we want to differentiate the minimum value of the free energy with respect to $T$. Below $T_c$ we have $f=f_0+a^3/6b^2$ and hence we deduce that $T\partial^2 f/\partial T^2\propto t^{3n-2}$ and thus $\alpha=2-3n$. Note that above $T_c$ the free energy is constant and hence the specific heat vanishes. To determine $\delta$ we add $-m h$ to the free energy where $h$ is
a background source. We now have $\partial f/\partial m=am +b m^2- h$ and at $T=T_c$, where $a=0$, we deduce that the equilibrium configuration has 
$m\propto h^{1/2}$ and hence $\delta=2$. Finally, we consider the susceptibility $\chi=\partial m/\partial h$. At equilibrium we have $am +b m^2- h=0$ and differentiating we deduce that
$\chi=1/(a+2 bm)$. For $T>T_c$ we have $m=0$ and 
$\chi=t^{-n}$, while for $T<T_c$ we have $m=-a/b$ and hence 
$\chi=-t^{-n}$. We thus deduce that $\gamma=n$. When $n=1$ the critical exponents
are thus given by
$(\alpha,\beta,\gamma,\delta)=(-1,1,1,2)$, exactly as we saw in our holographic phase
transition. 

We now return to the issue of the global instability for $m<0$. 
We first note that the instability would be
eliminated if we were restricted to configurations with $m\ge 0$. Interestingly, the critical exponents that we have obtained were discussed in the context of a continuum generalisation of the Ashkin-Teller-Potts models associated with
percolation problems, by imposing such a restriction \cite{PhysRevB.13.4159}. 
Note that we have no restrictions on the sign of the expectation value $\langle {\cal O}_\psi\rangle$, so
this perspective is not available for our holographic phase transition. 

It is also worth pointing
out that if we try to stabilise the free energy with higher powers of $m$, a quartic for example, then the model
has a first order transition, again unlike what we see in our holographic transition.
More explicitly we can add a term $cm^4/4$ to the free energy in 
\eqref{felg} with $c>0$. Now for high temperatures, $a> b^2/(4c)$, the free energy has a minimum at $m=0$. For $2b^2/(9c)<a<b^2/(4c)$ there is an additional minimum at $m=m_1\equiv -b/(2c)-[b^2-4 a c]^{1/2}/(2c)$, which, has higher free energy than
the minimum at $m=0$. For $0<a<2b^2/(9c)$ the minimum at $m_1$ has lower free energy than the minimum at $m=0$ and there is a first order transition at $a=2b^2/(9c)$. For $a<0$, $m=0$ becomes a maximum of the free energy with a new minimum 
appearing at
$m=m_2\equiv -b/(2c)+[b^2-4 a c]^{1/2}/(2c)$. This $m_2$ minimum is the one associated with the critical exponents for the cubic with $c=0$ that we discussed above, but it is simple to see that the $m_1$ minimum is always preferred.

In summary, we see that while the cubic Landau-Ginzburg model for a single scalar order parameter in a certain sense gives
rise to the critical exponents we see in our holographic phase transition it
does not capture key features. Perhaps a model containing more fields might be more effective.

\providecommand{\href}[2]{#2}\begingroup\raggedright\endgroup
\end{document}